\documentstyle[aps,multicol]{revtex}
\draft 
\title{Magnetic Force Exerted by the Aharonov-Bohm Line}
\author{Andrei Shelankov $^{*}$}
\address{
 Department of Theoretical Physics, Ume{\aa} University, 901 87
Ume{\aa}, Sweden 
}
\date{ cond-mat/9802158: 14 February 1998, revised 2 April 1998} 

\begin{document}
\maketitle

\begin{abstract}
The problem of the scattering of a charge by the Aharonov-Bohm (AB)
flux line is reconsidered in terms of finite width beams.  It is shown
that despite the left-right symmetry in the AB scattering
cross-section, the charge is deflected by the AB-line as if by the
``Lorentz'' force.  The asymmetry originates from almost forward
scattering within the angular size of the incident wave.  In the
paraxial approximation, the real space solution to the scattering
problem of a beam is found as well as the scattering $S$-matrix.  The
Boltzmann kinetics and the Landau quantization in a random AB array are
considered.
\end{abstract}


\begin{multicols}{2}

The Aharonov-Bohm (AB) flux line is an idealized construction
originally designed to discuss the role of the vector potential in
quantum mechanics \cite{AhaBoh59}. The magnetic field around the
AB-line is zero but the gauge vector potential is finite being
generated by the magnetic flux $\Phi $ concentrated in the line.
Nowadays, the AB-line is very popular in various contexts: As a
carrier of Chern-Simon's field, the AB-line attached to a particle
allows one to build two-dimensional objects obeying a fractional
statistics (anyons), or composite fermions in the theory of strongly
correlated electronic systems (Quantum Hall Effect or high-T$_{c}$
oxides).  In many respects, vortex lines in superfluids are similar to
AB-lines: The superflow around a vortex line in $He-II$ acts on a
normal excitation like the vector potential of an AB-line on a charge.
Problems of the vortex dynamics, e.g the existence of the Iordanskii
force \cite{Ior65}, are currently under active debate
\cite{Magnusnia97,ThoAoNiu97,Son97}.

There exists a vast literature on a charge interacting with an AB-line
(for a review see \cite{OlaPop85}).  Surprisingly, there is a question
which is still controversial.  One can formulate the question as: Does
the AB-line exert a Lorentz-like magnetic force on a moving charge?
Or, in other words, whether the charge is deflected right-left
asymmetrically revealing the absence of the mirror symmetry (${\cal
P}$) broken by the magnetic flux.  Despite the fact that the exact
solution to the AB-line scattering problem has been known since the
original paper of Aharonov-Bohm \cite{AhaBoh59}, there is no common
opinion on the subject.  Different people give opposite answers
saying:

\noindent 
{\it No:}
Any effect of scattering can be expressed via the differential
cross-section $d\sigma /d \varphi$ , $\varphi $ being the scattering
angle.  The cross-section known from \cite{AhaBoh59},
\begin{equation}
{d \sigma \over d\varphi  }=
{\lambdabar  \over 2 \pi}\;
{\sin^{2}\pi \tilde{\Phi} \over \sin^{2}{\varphi \over 2 } } 
\label{bfa}
\end{equation}
is $\varphi \rightarrow - \varphi $ symmetric, and, therefore, there
is no left-right asymmetry in the outgoing wave ($\tilde{\Phi}= \Phi
/\Phi_{0}$, $\Phi_{0}= hc/e$, and $\lambdabar =1/k$ is the particle
wave length).  The net transverse momentum transfer must be zero
together with the Lorentz force.  The only way to reveal the broken
${\cal P}$ is via the interference of the AB- and potential scattering
as e.g. discussed in \cite{MarWil88}.

\noindent 
{\it Yes:}
Any effect allowed by symmetry is there.  The symmetry of the problem
is that of a charge in a magnetic field and the transverse force must
be finite.  As far as calculations are concerned, the force is given
by a divergent integral (over $\varphi $) or sum (over partial waves).
The divergence can be handled and the result seems to be reasonable:
it agrees with the derivation based on the momentum balance 
\cite{OlaPop85}.

The purpose of this paper is to reconsider the problem.  The main idea
is that the difficulties and ambiguities are due to the combination of
two factors: (i) the infinite range of the vector potential; of the
AB-line; and (ii) the infinite extension of the incident plane wave in
the Aharonov-Bohm solution, allowing absolute resolution in the
direction of propagation.  The plane wave solutions are not suitable
for extracting physics from them because of the forward scattering
singularity.  Instead, one should analyze beams with a finite angular
size.  Then, the singularity at $\varphi =0$ is regularized in a
natural way.

The paper is organized as follows.  First, to check the very existence
of the transverse ``Lorentz force'', I analyze the gedanken experiment
where an incoming beam of a finite effective width $W$ hits an AB-line
and the deflection of the beam measures the magnetic force exerted by
the line.  The calculations are done in paraxial approximation
\cite{parax}.  At the expense of fine details on the scale of the wave
length $\lambdabar$, the paraxial approximation allows one to find
rather easily the wave function for small scattering angles at
distances $\gg \lambdabar $.  As expected from the symmetry arguments,
a finite deflection is observed.  However, the deflection angle being
$\sim \lambdabar /W$ tends to zero when the incoming beam transforms
to an infinite plane wave, $W \rightarrow \infty $ (in agreement with
the argument based on the symmetry of the cross-section).  The
transverse momentum transfer is found for arbitrary incoming wave, and
the result is presented in terms of the effective force.  To
understand how the effect of many lines adds together, I consider a
random array of the AB-lines, a model similar to that of Desbois {\it
et al.}\cite{Des}.  On average, the anomalous asymmetric small-angle
scattering amounts to an effective magnetic field proportional to the
density of lines. The kinetic equation and the resistivity tensor, as
well as the Landau quantization are discussed. Implications of the
results are discussed at the end of the paper.

Consider the experiment where a particle of mass $m$, charge $e$ and
momentum $p= \hbar k = \hbar / \lambdabar $ moves on the $x-y$ plane
in the $x$-direction from $- \infty$ and meets the Aharonov-Bohm line
piercing the plane at $\bbox{r}=0$.  The distribution with respect to
the transverse coordinate $y$ is measured, and the averaged value
$\bar{y} (x)$ defines the ``trajectory'' from which the deflection of
the particle by the AB-line is extracted.  To make the transverse
coordinate meaningful, the stationary incident wave is beam-like with
a finite transverse size $W\gg \lambdabar $ (e.g. a wave having passed
through an aperture of the width $W$).

In the paraxial theory \cite{parax}, a particle moving at a small
angle to the $x-$axis is described by the wave function of the form
$\Psi = e^{ikx} \psi$, where 
$\psi (x,y) $ is slow, $ |\nabla \psi | \ll k |\psi|$.  Neglecting
${\partial^{2} \psi\over{\partial x^{2}}}$ $\ll k {\partial \psi
\over{\partial x}}$ in the stationary Schr\"odinger equation, one
comes to the paraxial equation:
\begin{equation}
i  v \partial_{x}\psi  = -{1\over{2m}}\partial^{2}_{y}\psi
\label{4ea}
\end{equation}
where the velocity $v = \hbar k/m $ and $\bbox{\partial}\equiv \hbar
\bbox{\nabla }-i {e\over{c}}\bbox{A}$, $\bbox{A}$ being the vector
potential chosen below as $A_{y}=0$ and $A_{x}= - \Phi \delta (x)
\,\theta (y)$.

The incoming wave, $\psi(x<0, y)$, is controlled by conditions of the
experiment such as screens, apertures {\em etc}.  Leaving the
preparation of the incoming wave out of the picture, $\psi(x= - 0, y)
\equiv \psi_{in}(y) $ can be taken as the input to the scattering
problem.  Solving Eq.(\ref{4ea}) in the immediate vicinity of the line
$x=0$, where the vector potential is concentrated in the chosen gauge,
one finds
\begin{equation}
\psi(+0,y)= \psi_{in}(y)\exp \big( - 2\pi i \tilde{\Phi }
\theta (y)\big) \ .
\label{5ea}
\end{equation}
Further propagation is free, and 
the outgoing wave is
\begin{equation}
\psi(x,y)= \int_{- \infty }^{\infty }
d y'\,
G_{0}(y-y';x)
\psi(+0,y') \;\; , \;\;  x>0 ,
\label{55ea}
\end{equation}
 where 
$
G_{0}(y;x)=  \theta (x)
\sqrt{{k\over  2\pi ix}}
 e^{{ik\over{2  x}} y^{2}}
$.
Or, using Eq.(\ref{5ea}),
\begin{eqnarray}
e^{i \pi \tilde{\Phi}}\psi (x,y)     & =  &  
\cos \pi \tilde{\Phi}\,\,\psi_{0}(x,y)            \nonumber\\   
     &  &  \hspace{-10ex} + i \sin \pi \tilde{\Phi} 
\int_{- \infty }^{\infty }
d y'\,
G_{0}(y-y';x) 
{y'\over |y'| }
\psi_{in}(y') \; , \;  x>0 ,
\label{hfa}
\end{eqnarray}
where $\psi_{0}(x,y)= 
\int_{- \infty }^{\infty }
d y'\,
G_{0}(y-y';x)
\psi_{in}(y')
$
is the solution in the absence of the line.

Given the incoming wave $\psi_{in}(y)$, Eqs.(\ref{55ea}) or
Eq.(\ref{hfa}) allows one to find the outgoing wave at $x \gg
\lambdabar $ in the small angle region $|\varphi | \ll 1$, $\varphi =
y/x$.  

For the {\em infinite} plane incident wave, {\it i.e.}
$\psi_{in}(y)=\psi_{0}(x,y)=1$, Eq.(\ref{hfa}) immediately gives (up
to a gauge transformation) the Aharonov-Bohm solution at $|\varphi
|\ll 1$ \cite{OlaPop85}, thus confirming the validity of the paraxial
approximation. The paraxial solution $\psi(x,y) $ depends on the
coordinates $x$ and $y$ only in the combination $s = y/
\sqrt{2\lambdabar x}= \varphi \sqrt{x/2 \lambdabar}$. (When $s$
varies, $\psi (s)$ defines a curve on the complex $\psi$-plane. One
can show that the curve is nothing but the well-known Cornu spiral
\cite{BorWol59}, somehow scaled and shifted; the curve length along
the spiral is proportional to $s$.)  At $|s|\gg 1$, {\it i.e.}
$|\varphi| \gg \sqrt{\lambdabar /x}$, the wave function acquire the
asymptotic form usual for a scattering problem; the corresponding
scattering amplitude $\propto 1/ \varphi $ giving raise to the AB
cross-section in Eq.(\ref{bfa}) ($\varphi \ll 1$).  In the forward
direction $y=s=0$ the second term in the r.h.s. of Eq.(\ref{hfa})
vanishes, and, in agreement with \cite{OlaPop85,BerChaLar80},
$|\psi|^{2}= \cos^{2}\pi \tilde{\Phi}$, whatever distance from the
AB-line.  The anomalous behaviour when the wave function differs from
the incident wave and at the same time does not depend on the distance
to the scatterer, takes place at $|s|\lesssim 1$ in the progressively
narrow angle range $|\varphi| \lesssim \sqrt{\lambdabar /x}$. This is
the singularity which causes calculational problems in the standard
partial wave analysis of the scattering theory.  Obviously, the
singularity is absent when the direction $\varphi =0$ is ``blurred''
as it is the case if the incident beam is of a finite angular size.
When regularized, the forward scattering anomaly turns out to be
responsible for the asymmetry in the AB-scattering as it is shown
below.

A {\em finite} width beam $\psi_{in}(y)= \exp ( - |y|/W )$,  generates at $
x \gg W^{2}/\lambdabar$ the following angular distribution,
$P(\varphi)$, in the  outgoing wave
($P(\varphi ) d \varphi = x |\psi|^{2} d \varphi $):
\begin{equation}
P(\varphi )= 
{2 \lambdabar  \over \pi   } 
{(\varphi \,\sin \pi \tilde{\Phi} -\varphi_{0}\cos \pi \tilde{\Phi})^{2}\over (\varphi^{2}+ \varphi_{0}^{2})^{2}}
\;\;
, \;\;  x \gg W^{2}/\lambdabar  \; ,
\label{ffa}
\end{equation}
$\varphi_{0} = \lambdabar / W \ll 1 $ being the beam angular width.
As expected, the angular distribution is regular at $\varphi =0$.
Importantly, if $\varphi_{0}\neq 0$, the distribution is {\em
asymmetric} at the angles, $|\varphi |\lesssim \varphi_{0}$.  To
quantify the asymmetry in general case, I calculate below the integral
effect {\it i.e.}  the deflection of the beam as a whole (``the
trajectory bending'').

The transverse position of the particle at a given $x$ is defined as $
\bar{y} = \int_{- \infty }^{\infty } dy\, y |\psi(x,y)|^{2}$, and the
angle of propagation is $\bar{\varphi} =d \bar{y}/ dx$.  It follows
from Eq.(\ref{4ea}) (analogous to the Ehrenfest theorem) that $d
\bar{y}/ dx = \langle \hat{p}_{y}\rangle/mv$, $ \hat{p}_{y}$ being the
kinematical momentum.  In the chosen gauge, $\langle
\hat{p}_{y}\rangle_{out}= \int_{-\infty }^{\infty }dy \,
\psi^{*}(x,y){\hbar \over i} {\partial\over{\partial y}}\psi(x,y)$
with $\psi (x,y)$ given by Eq.(\ref{55ea}).  In the force free region,
$\langle \hat{p}_{y}\rangle_{out}$ does not depend on $x$, and the
integral can be conveniently calculated at $x \rightarrow + 0$ using
Eq.(\ref{5ea}).  The deflection angle $\Delta \varphi \equiv
\bar{\varphi}_{out}-\bar{\varphi}_{in}$ is $\Delta \varphi = { \Delta
p_{y}/ p }$ where $\Delta p_{y}= \langle \hat{p_{y}} \rangle_{out} -
\langle \hat{p_{y}} \rangle_{in} $, After simple calculation
\cite{foot},
\begin{equation}
\Delta p_{y} = - \hbar  |\psi_{\text{in}}(0)|_{N}^{2} \sin 2\pi
\tilde{\Phi} \ ,
\label{ida}
\end{equation}
where 
$ |\psi_{\text{in}}(0)|_{N}^{2} =  |\psi_{\text{in}}(0)|^{2}/
\left(\int_{-\infty }^{\infty }dy |\psi_{in}|^{2}\right)$.

We see that indeed the AB-line deflects particles asymmetrically, with
the left-right asymmetry $\Delta \varphi = \Delta p_{y}/p $ controlled
by parity-odd $\Phi $.  One could naively expect the asymmetry
thinking in terms of the classical Lorentz force exerted by the
magnetic field ``inside'' the line.  However, in contrast to the
classical expectations, $\Delta \varphi $ is {\em not} proportional to
the field strength but is periodic in $\Phi $, revealing a quantum
origin of the effect.  In agreement with general arguments
\cite{OlaPop85}, the deflection Eq.(\ref{ida}) is finite only if
$\psi_{in}(0)\neq 0$, {\it i.e.}  the incoming wave must overlap with
the line.

From the estimate $W|\psi_{\text{in}}(0)|_{N}^{2}\sim 1$ and
Eq.(\ref{ida}), deflection $\Delta \varphi $ is of order and never
exceeds the beam angular size $\varphi_{0} \sim \lambdabar / W$,
supporting the conclusion made from Eq.(\ref{ffa}).  This means that
the transverse effect cannot be discussed in terms of the scattering
amplitude: Within the cone $|\varphi | \sim \varphi_{0}$, the
scattered and incoming wave inevitably overlap and the split of the
wave into incoming and scattered components is rather arbitrary.
Besides, the forward scattering is singular and, as was discussed by
Berry {\it et al.}  \cite{BerChaLar80} one should deal with the full
wave function, which is regular in the forward direction, rather than
split it into two singular parts.

Instead, one may use the description in terms of the $S-$matrix:
$F_{out}(\varphi)= \int_{-\infty}^{\infty} d \varphi' S(\varphi ,
\varphi')F_{in}(\varphi')$, with the amplitudes $F$'s defined in the
momentum representation,
\begin{equation}
\psi = \int_{-\infty}^{\infty} d \varphi
F_{in/out}(\varphi ) e^{ik \varphi y - i {1\over{2}} kx \varphi^{2}}
 ,\,x<0/x>0.
\label{xia}
\end{equation}
It follows from Eq.(\ref{5ea}), that $S(\varphi , \varphi')$ is the
Fourier transform of 
$$S(y,y')= 
\delta (y-y')\exp( - 2\pi i \tilde{\Phi }\theta (y))
\cdot \exp(  \delta |y'|) $$
where the last factor with a regularization parameter $\delta = + 0$
is introduced with the understanding that the incident wave has a
finite extension in the $y$-direction.  Finally, $S$-matrix reads
\begin{equation}
S(\varphi , \varphi')= \delta (\varphi -\varphi ')+ {1\over \pi} \sin
\pi \tilde{\Phi} {e^{-i \pi  \tilde{\Phi}}\over \varphi' - \varphi + i
  \delta }\; \;\; , \;\; \delta = + 0 \; .
\label{yia}
\end{equation}
$S(\varphi , \varphi')$ is unitary, its overall phase is
gauge-dependent. 

 For an incident wave with physically reasonable
$F_{in}(\varphi)$, this regularized expression for the ${S}-$matrix
gives the outgoing wave $F_{out}(\varphi)= \hat{S}F_{in}(\varphi)$
finite and smooth at any angle:
\begin{equation}
e^{i \pi \tilde{\Phi}}\,
F_{out}(\varphi )= \cos \pi \tilde{\Phi}\,
F_{in}(\varphi )+ {1\over \pi}\sin \pi \tilde{\Phi}\,
\rlap{\hspace{.27em}$-$}\int d \varphi ' \,{F_{in}(\varphi')\over \varphi'
  -\varphi  }\, ,
\label{pja}
\end{equation}
where $\rlap{\hspace{0em}$-$}\int$ denotes the principal value of the
integral. Coming from the $\delta -$function contributions in the
$S-$matrix, the first term in the r.h.s.  gives
a modified incident wave, attenuated and with a phase shift (gauge dependent).
The second term  reproduces 
the  AB-scattering in the limit of small scattering angles.

It is clear now that the left-right asymmetry in the outgoing
intensity $|F_{out}|^{2}$ may come only from the interference of the
two terms in Eq.(\ref{pja}): The cross product is the only contribution to
$|F_{out}|^{2}$ with the magnetic symmetry, {\it i.e.}  odd relative
to $\Phi \rightarrow -\Phi$.  Obviously, the magnetic interference
piece is present only at the angles where $F_{in}(\varphi)\neq 0$ {\it
i.e.}  of order of the angular width of the incident wave $|\varphi
|\sim \lambdabar /W$ .  Loosely speaking, the asymmetry and the
magnetic force exerted by the AB-line originates in the interference
of the scattered and incident waves.

Qualitatively similar physics was conjectured in
Ref.\cite{NieHed95}. Importance of ``auto-interference'' was emphasized
in Ref. \cite{Ste95} in the context of the time-dependent problem of
scattering of a wave packet by the Aharonov-Bohm line.

The AB-scattering is an interesting example when the forward
scattering singularity is not just a nuisance, as in the Coulomb
scattering case, but it is totally responsible for a qualitative
effect -- the asymmetry in the scattering.

In Eq.(\ref{ida}), $\Delta p_{y}$ is the transverse momentum transfer
per collision. Multiplying it by the collision rate $\dot N $, one
gets a combination, ${\cal F}_{y}= \Delta p_{y} \dot N$, which has the
meaning of the force acting on the charge from the line.  The
collision rate is found as $\dot N = \int_{-\infty }^{\infty }dy
j_{x}(x,y)$, $j_{x}$ being the current density.  In the paraxial
approximation $j_{x}= v |\psi |^{2}$, and using Eq.(\ref{ida}) the
force ${\cal F}_{y}= - \hbar v |\psi (0) |^{2}$.  In terms of the full
wave function $\Psi_{in}(x,y)= e^{ikx}\psi_{in}(x,y)$ the effective
``Lorentz force'' reads in the vector form as
\begin{equation}
 \bbox{{\cal F}}_{L} =  
\hbar \sin 2\pi \tilde{\Phi}\;
\left(\bbox{J}_{in}
\bbox{\times e  }_{z}\right)  
\label{ifa}
\end{equation}
where $\bbox{J}_{in}$ is the current density in the {\it incident
wave} at the position of the line: 
$\bbox{J}_{in}= 
{\hbar \over m}\ 
\Psi^{*}_{in}\bbox{\nabla }\Psi_{in} |_{\bbox{r}=0}$.  

Eq.(\ref{ifa}) is in agreement with earlier results
\cite{Cle68,NieHed95,Son97} where the force was extracted from
(divergent) sums in the partial wave analysis.  In passing, if one
uses wave packets built from the exact AB-solutions, the sums become
converging and the doubts in \cite{ThoAoNiu97} concerning results of
\cite{Son97,Cle68,NieHed95} can be readily rejected.

The deflection in Eq.(\ref{ida}) or the force in Eq.(\ref{ifa}) are
dependent on the details of the wave-packet.  A more simple picture
emerges when the charge sees many lines and the structure details tend
to average out.  I consider a model of an AB-array where the position
of a line and its flux are random; the density of the array is
$d_{AB}$.  After averaging with respect to the randomness in the
array, one comes to the Boltzmann-type equation for the distribution
function $n_{\varphi }(p,\bbox{r})$; $p= |\bbox{p}|$ and $\varphi $
shows the orientations of the particle momentum $\bbox{p}$.  The
Boltzmann equation, the derivation \cite{foot1} of which will be
presented elsewhere, reads
\begin{equation}
\bbox{v\!\cdot\! \nabla }n_{\varphi } 
 - {e \tilde{B}\over mc } {\partial n_{\varphi}\over{\partial \varphi }} +
{1\over 2 \tau_{_{AB}}} \int_{0}^{2\pi }\!
 {d \varphi' \over 2 \pi  } {\left( n_{\varphi }- n_{\varphi'}\right)\over \sin^{2}{{\varphi -\varphi ' \over 2}}} 
=0
\label{nnea}
\end{equation}
where 
\begin{equation}
{1\over \tau_{_{AB}}} =  
{2\hbar \over m }
d_{_{\text{AB}}} 
\big<\sin^{2}\pi \tilde{\Phi} \big> \; , \;  
\tilde{B}=  d_{_{\text{AB}}} {\Phi_{0}\over 2\pi } 
\big< \sin 2 \pi  \tilde{\Phi} \big>,
\label{qqea}
\end{equation}
here $<\ldots > $ denotes the averaging over the probability
distribution of the flux of a line.

The structure of the kinetic equation Eq.(\ref{nnea}) is rather
obvious.  The collision integral part corresponds to the picture where
the AB-lines play the role of impurities independently scattering
particles in accordance with the cross-section in Eq.(\ref{bfa}).  The
divergence of the scattering rate at $\phi \rightarrow 0$ is
unimportant for kinetics as long as the {\it transport} scattering
time ( $=\tau_{AB}$) is finite.  Accounting for the left-right
asymmetry in the forward scattering, $\tilde{B}$ in Eq.(\ref{nnea})
enters like a magnetic field.  The bending of the trajectories and the
expression for $\tilde{B}$ Eq.(\ref{qqea}) can be understood from
Eq.(\ref{ida}) or Eq.(\ref{ifa}) if one sums up the transverse
momentum acquired in sequential multiple collisions.

If typically $\Phi \ll \Phi_{0}$, then $\tilde{B}\approx B_{z}$, where
$B_{z}= d_{AB} \langle \Phi \rangle $ is the macroscopic magnetic flux
density. In this limit, the AB-array is equivalent to Gaussian
$\delta-$correlated random magnetic field.  If $\Phi \sim \Phi_{0}$,
the behaviour is essentially non-Gaussian: Unlike induction $B_{z}$,
the effective field is a periodic function of the flux.  As expected,
integer and half-integer \cite{MarWil88,NieHed95} fluxes do not
contribute to $\tilde{B}$.  For a general distribution in $\Phi$,
$\tilde{B}$ and $B_{z}$ may be in any relation.  Remarkably,
$\tilde{B}$ and the Lorentz force may be finite even if macroscopic
induction $B_{z}=0$ and, ${\cal P}$ and time-reversal symmetries are
macroscopically preserved.

Seeing that Eq.(\ref{nnea}) has the standard form and $1/ \tau_{AB}$
and $\tilde{B}$ are energy independent, one can immediately write down
the (Drude) resistivity tensor: $\rho_{xx}^{-1}= e^{2} N_{0}
\tau_{AB}/ m$, $N_{0}$ being the particle density, and $\rho_{yx}=
(\Omega_{c}\tau_{_{AB}})\rho_{xx}$ where $\Omega_{c}= e \tilde{B}/mc$
(see also \cite{NieHed95,Des}).  For the Hall angle $ \theta_{H}$, one
gets 
$\tan\theta_{H} =
\Omega_{c} \tau_{AB} = 
< \sin 2\pi \tilde{\Phi}>/ 2 <\sin^{2}\pi \tilde{\Phi}>$.  
In particular,
\begin{equation}
\tan\theta_{H} = \cot \pi  \tilde{\Phi}
\label{rrea}
\end{equation}
if the lines have same flux $\tilde{\Phi}$. 
 
The large $\theta_{H}$ when $\tilde{\Phi}\ll 1$, indicates that the
particles move along the Larmor circles.  The Landau-type quantization
of the periodic motion is then expected.  In the semiclassical
approximation, the oscillations in the density of states $\delta
\rho_{osc}(E) \propto e^{-\gamma }\cos (2\pi E/\hbar \Omega_{c})$ in a
random Gaussian magnetic field have been considered by Aronov {\it et
al.}  \cite{AroAltMir95} (see also Desbois {\it et al.}  \cite{Des}).
Adjusting \cite{AroAltMir95}, the damping parameter reads
\begin{equation}
  \gamma =
{2\pi \over \Omega_{c} \tau_{_{AB}} }\, {E\over \hbar \Omega_{c} } =
  {\pi \over 2 \lambdabar^{2} d_{AB}}\;\; . 
\label{2ea}
\end{equation}
Note that in the semiclassical situation, $E \gg \hbar \Omega_{c}$,
the damping is strong even when $\Omega_{c}\tau_{AB}> 1$ and the
periodic Larmor orbiting is well pronounced.  Specific to the random
magnetic field environment, the damping Eq.(\ref{2ea}) increases with
the energy $E$: Similar to inhomogeneous broadening mechanism, the
damping here is due to the fluctuations in the number of the AB- lines
threading the Larmor circles, the area of which increases $\propto E$.

In conclusion, I have reconsidered the problem of the scattering by
the gauge vector field of the Aharonov-Bohm line.  It has been
demonstrated that the difficulties in the previous calculations,
ultimately related to the infinite range of the vector potential, can
be avoided if one deals with a beam of a finite transverse size $W$
rather than a infinitely extended plane wave.  As allowed by the low
symmetry of the problem, the moving charge is asymmetrically deflected
by the line (provided the incident wave overlaps with it) despite the
fact that the Aharonov-Bohm scattering-cross section is symmetric. The
asymmetry in the angular distribution may be attributed to the
interference of the incident and scattered waves, and it exists only
at the scattering angles of order of the angular width of the incident
wave $\varphi_{0}\sim \lambdabar /W$.  The derivation is done in the
framework of the paraxial approximation which proves to be very
efficient.  The magnetic ``Lorentz force'', i.e. the transverse
momentum transfer, has been calculated.  Earlier results in
\cite{Cle68,Son75,NieHed95,Son97} are confirmed.  For an arbitrary
incoming wave, the ``Lorentz'' force Eq.(\ref{ifa}) acting on the
charge is expressed via the current density in the incident wave at
the position of the line. 

The AB-problem solves also the problem of the scattering of a phonon
by the superflow around vortex line in $He-II$.  The superflow is not
a gauge-field, and the equivalence holds only in the lowest
approximation with respect to the vortex circulation $\kappa $.  The
``Lorentz force'' $\propto \Phi$ acting on the charge translates
\cite{Son75} as the (minus) force $\propto\kappa $ acting on the
vortex line, the Iordanskii force \cite{Ior65}.  The very existence of
the Iordanskii force has been recently doubted on the grounds of
hardly verifiable general arguments \cite{Magnusnia97} as well as
claiming \cite{ThoAoNiu97} a technical mistake in the previous works
on the subject.  The present calculations, which are free from
divergences and ambiguities of some earlier papers, support the
existing understanding on the Iordanskii force in Helium-II
\cite{Ior65,Son75,Son97} and its role in the vortex dynamics.

Finally, a random array of AB-lines has been considered.  The
resistivity tensor as well as Landau quantization in the array has
been discussed.  The array with typical flux in a line $\sim \Phi_{0}$
gives an example of random magnetic field system with non-Gaussian
fluctuations.  An interesting observation here is that the effective
Lorentz force seen by a charged particle in the array as well as the
Landau quantization may persist even when the macroscopic flux density
is zero.

I am grateful to S. Levit, A. Mirlin, L. Pitaevskii, and P. W\"olfle
for discussions and also to E. Sonin for critical remarks.  The study
began during my stay at the Institut f\"ur Theorie der Kondensierten
Materie, Universit\"at Karlsruhe, and I would like to thank all the
the members for their hospitality.  This work was supported by SFB 195
der Deutsche Forschungsgemeinschaft and in part the Swedish Natural
Science Research Council.

\end{multicols}


\begin{references}
\bibitem{foot} 
Because of the discontinuity of
$\psi(+0,y)$ in Eq.(\ref{5ea}) at $y=0$, which is an artifact of the
paraxial approximation, some caution is needed.  The correct limiting
procedure, where details of the behaviour at $y=0$ are unimportant, is
the following: $\partial \psi /\partial y $ is substituted for
$\left(\psi (y + a , x)- \psi (y -a , x) \right)/2a $ and the limit $a
\rightarrow 0$ is taken {\em after} the $y$-integration at $x = + 0$.
\end{references}
\end{document}